\documentclass[fleqn,twoside]{article}
\usepackage{epsf,multicol,ifthen}
\usepackage{ujp}
\usepackage[cp1251]{inputenc}
\usepackage[english,ukrainian]{babel}
\usepackage{amstext}
\usepackage{amssymb}
\usepackage{amsmath}
\usepackage{cite}
\usepackage {graphicx}

\nazva{OCCURRENCE OF PAIRWISE ENERGY LEVEL DEGENERACIES IN
\boldmath{$q,p$}-OSCILLATOR MODEL }

\udk{541.183}

\nazvacol{PAIRWISE ENERGY LEVEL DEGENERACIES IN $q,p$-OSCILLATOR }%
\razd{GENERAL QUESTIONS OF THERMODYNAMICS, STATISTICAL\\ PHYSICS,
AND QUANTUM MECHANICS}%
\avtor{A.M. GAVRILIK, A.P. REBESH}
\avtorcol{A.M. GAVRILIK,
A.P.REBESH}
\inst{Bogolyubov Institute for Theoretical Physics, Nat. Acad. Sci. of Ukraine}%
\adr{(14b, Metrolohichna Str., Kyiv 03680, Ukraine; e-mail: omgavr@bitp.kiev.ua)}%

\begin{document}
\setcounter{page}{586}
 \maketitl
\begin{multicols}{2}

\anot{It is demonstrated that a two-parameter
deformed oscillator with the deformation parameters $q,p$
such that $0<q,p\le 1$   %belonging to the intervals
exhibits the property of ``accidental'' two-fold (pairwise) energy
level degeneracy of the classes $E_m=E_{m+1}$ and $E_0=E_{m}$. The
most general case of degeneracy of $q,p$\,-oscillators of the form
$E_{m+k}=E_m$ (with $k\ge 1$ for $m\ge 1$ or $k\ge 2$ for $m=0$)
is briefly discussed.}

\section{Introduction}

The so-called $q$-deformed oscillators ($q$-oscillators) remain,
from their appearance till now, a very popular subject of
investigations including their diverse applications (see, e.g.,
\cite{Zhang,Bona} and references therein). Much attention has been
paid to the two most distinguished versions of $q$-oscillators:
the one proposed by Biedenharn and Macfarlane \cite{Bied,Mcf} (BM
$q$-oscillator) and the other introduced by Arik and Coon
\cite{AC} (AC $q$-oscillator).

It is well known that, unlike the AC $q$-oscillator, the BM
version admits not only real but also phase-like complex values of
the deformation parameter $q$. Such a distinction leads to
essentially differing aspects of their particular applications.
Let us note that, among others, there is the well-known property
of the BM $q$-oscillator consisting in a possibility of certain
degeneracies and periodicity appearing in case of $q$ being a root
of unity, the most popular values for the BM-type $q$-oscillator.
Say, for $q=\exp(\frac{{\rm i}\pi} {2n+2})$, the following two
neighboring energy levels coincide: $E_{n+1}=E_{n}$. This equality
along with other coincidences leads to a kind of periodicity and
naturally makes the corresponding phase space both discrete and
finite~\cite{Bonats}.%\looseness=1

One can then wonder whether some kind of 'accidental' degeneracy
(occurring without any obvious underlying symmetry) can be a
peculiar feature of the AC $q$-oscillator, and the answer is
negative: the only possible case requires the value $q=0$, but
usually this value is excluded from the treatment.

The latter conclusion is however not the ultimate
statement concerning $q$-deformed oscillators and,
as recently shown \cite{GR-1}, yet another version of
$q$-oscillator which has been termed
the ``Tamm-Dancoff cutoff'' deformed oscillator in \cite{Odaka, Jagan}
and does possess the property of ``accidental'' degeneracy of the kinds
$E_m=E_{m+1}$, $E_0=E_{m}$, and some others.

The goal of the present paper is to analyze the analogous question
about possible 'accidental' degeneracies if one deals with more
general two-parameter (or $q,p$\,-)deformed oscillators. The
$q,p$\,-deformed oscillators introduced in \cite{Chakr-Jag} more
than 15 years ago provide the valuable and perspective tools for
obtaining nonstandard $q$-oscillators and for elaborating diverse
applications. It suffices to mention only a few following ones.

First, the $q,p$\,-deformed oscillators turn out to be rather
effective \cite{Kibl} in the phenomenological description of the
rotational spectra of (super)deformed nuclei.

Second, the concept of $q,p$\,-deformation, unlike the standard
harmonic oscillator, allows to account for more involved
reasons/aspects of the extension of the standard oscillator: for
the situation where the included interaction is highly nonlinear
(non-polynomial, with inclusion of all anharmonisms) and/or
employs the momentum operator; it may also involve the
non-constant position-dependent mass of the quantum-mechanical
particle \cite{Mizrahi}.

Third, the application elaborated in \cite{SIGMA} incorporates the
appropriate set of $q,p$\,-deformed oscillators ($q,p$\,-bosons)
for developing
the corresponding $q,p$\,-Bose gas model. Such a model is based on %exploits
the analytical expressions for the intercepts (strengths) of
the general $n$-particle momentum correlation functions obtained for
the first time in explicit form in \cite{AdGa} (note that these
results generalize the previously known formulas for two-particle
correlations in the AC and BM versions of the $q$-Bose gas model). As
such, the mentioned results were analyzed \cite{SIGMA} in the
context of their direct relevance to experimental data on the 2-
and 3-pion correlations collected during the RHIC/STAR and
CERN/SPS runs of relativistic heavy ion collisions.%\looseness=1

The paper is organized as follows.  In Section~2, we recall the
main facts about the phenomenon of 'accidental' double degeneracy
of energy levels within two particular (BM or TD) versions of
$q$-oscillators, respectively for $q$ being a root of unity or $q$
being real. The peculiarities of an analogous sort of degeneracies
manifested by the two-parameter (or the $q,p$\,-) deformed
oscillators are disclosed and explained in Section 3, where we
formulate, prove, and illustrate our basic statements. Section~4
is devoted to concluding remarks.

\section{``Accidental'' Degeneracies of \boldmath{$q$}-Oscillators}

\noindent To explain the idea of 'accidental' degeneracies of
energy levels, consider first the famous Biedenharn--Macfarlane
(or BM) $q$-oscillator [3, 4], whose defining relations are
%1
\begin{equation}
 a a^\dagger - q a^\dagger a = q^{-N} , \qquad
  a a^\dagger - q^{-1} a^\dagger a = q^{N} ,
\end{equation}
%2
\begin{equation}
[{ N},a]=- a    ,     \qquad [{ N},a^\dagger]= a^\dagger.
\end{equation}
Then, $a^\dagger a=[N]_q$ and $a a^\dagger=[N+1]_q$ where the
$q$-bracket reads
%3
\begin{equation}
 [X]_q \equiv \frac{q^X-q^{-X}}{q-q^{-1}} , \qquad  \quad
 [X]_q \ \xrightarrow{q\to 1} X.
\end{equation}
The Hamiltonian of the BM $q$-oscillator is taken to be
\[
H  = \frac{\hslash\omega}{2} (a a^\dagger + a^\dagger a).
\]
For convenience, we put ${\hslash\omega}=1$ in what follows. Using
the $q$-Fock space and its vacuum state $|0 \rangle$ such that
\[
a |0 \rangle = 0 \ , \quad \quad |n \rangle = \frac{(a^\dagger
)^n}{\sqrt{[n]_q!} } |0 \rangle  , \quad \quad N |n \rangle = n~|n
\rangle  ,
\]
where $[n]_q!=[n]_q[n\!-\!1]_q ... [2]_q [1]_q$, $[1]_q=1$,
$[0]_q=1$, the creation/annihilation operators act by the formulas
\[
a \ |n \rangle = \sqrt{[n]_q} \ |n-1 \rangle ,  \quad a^\dagger |n
\rangle = \sqrt{[n+1]_q} \ |n+1 \rangle .
\]
The spectrum $H|n\rangle = E_n |n\rangle$ of the Hamiltonian reads
%4
\begin{equation}
E_n =  \frac12 \Bigl( [n+1]_q  +  [n]_q \Bigr) .
\end{equation}
If $q\to 1$, \ $E_n = n + \frac12$; also $E_0 = \frac12$ for any
value of $q$.

For real $q\ne 1$, the {\em spectrum is not equidistant}. The most
interesting situation arises for phase-like $q$, $q=\exp(i\theta)$.

\subsection*{2.1. Level degeneracy of a \boldmath$q$-oscillator with \boldmath{$q$$=$${\rm e}^{i\theta}$}}

In the next two statements,  $n$ is any positive integer.

{\bf Proposition 1}.
%%%%%%%%%%%%%%%%%%%%%%%%%%%%%%%%%%%%%%%%%%%%

(i) Fix the angle $\theta$ to be
\[\theta=\frac{\pi (2k+1)}{2n+2} \quad {\rm with} \quad
k=0,\pm1,\pm2,...
\]
Then Eq. (4) yields
%\[
$E_{n+1}-E_{n}=\cos\frac{(2n+2)\theta}{2}, $
%\]
and, with the indicated $\theta$, the degeneracy $E_{n+1}= E_{n}$
follows,

(ii) Fix the angle $\theta$ to be
\[
\theta=\frac{\pi (2k+1)}{2n+3} \quad {\rm where} \quad
k=0,\pm1,\pm2,\ldots
\]
Then Eq. (4) yields $E_{n+2}-E_{n}=2\cos\frac{(2n+3)\theta}{2}
\cos\frac{\theta}{2}$, and, with this $\theta$, the degeneracy
$E_{n+2}= E_{n}$ follows.

This statement can be generalized as follows.

{\bf Proposition 2}. For $r\!\ge\!1$, let us fix the angle
$\theta$ as
%5
\begin{equation}
\theta=\frac{\pi (2k+1)}{2n+r+1}  \quad  {\rm with}  \quad
k=0,\pm1,\pm2,\ldots
\end{equation}
Then Eq. (4) yields the equality
%6
\begin{equation}
E_{n+r}\!-E_{n}=2\frac{\sin(r\theta/2)}{\sin(\theta)}\cos\frac{(2n+\!1\!+r)\theta}{2}
\cos(\theta/2),
\end{equation}
from which
the degeneracy
 %7
\begin{equation}
E_{n+r}= E_{n} , \qquad   r\ge1
\end{equation}
follows for the values of $\theta$ given in (5).

The indicated degeneracies, for $q=\exp({\rm i} \theta)$ being the
corresponding root of unity with $\theta$ a rational fraction of
$\pi$, lead to such consequences as periodicity, discreteness, and
finiteness [6] of the phase space of a BM-type $q$-oscillator.

%%%%%%%%%%%%%%%%%%%%%%%%%%%%%%%% figs. 1 %%%%%%%%%%%%%%%%%%%%%%%%%%%%
\begin{center}\noindent
\includegraphics[width=8.7cm]
%%%{e67.eps}
{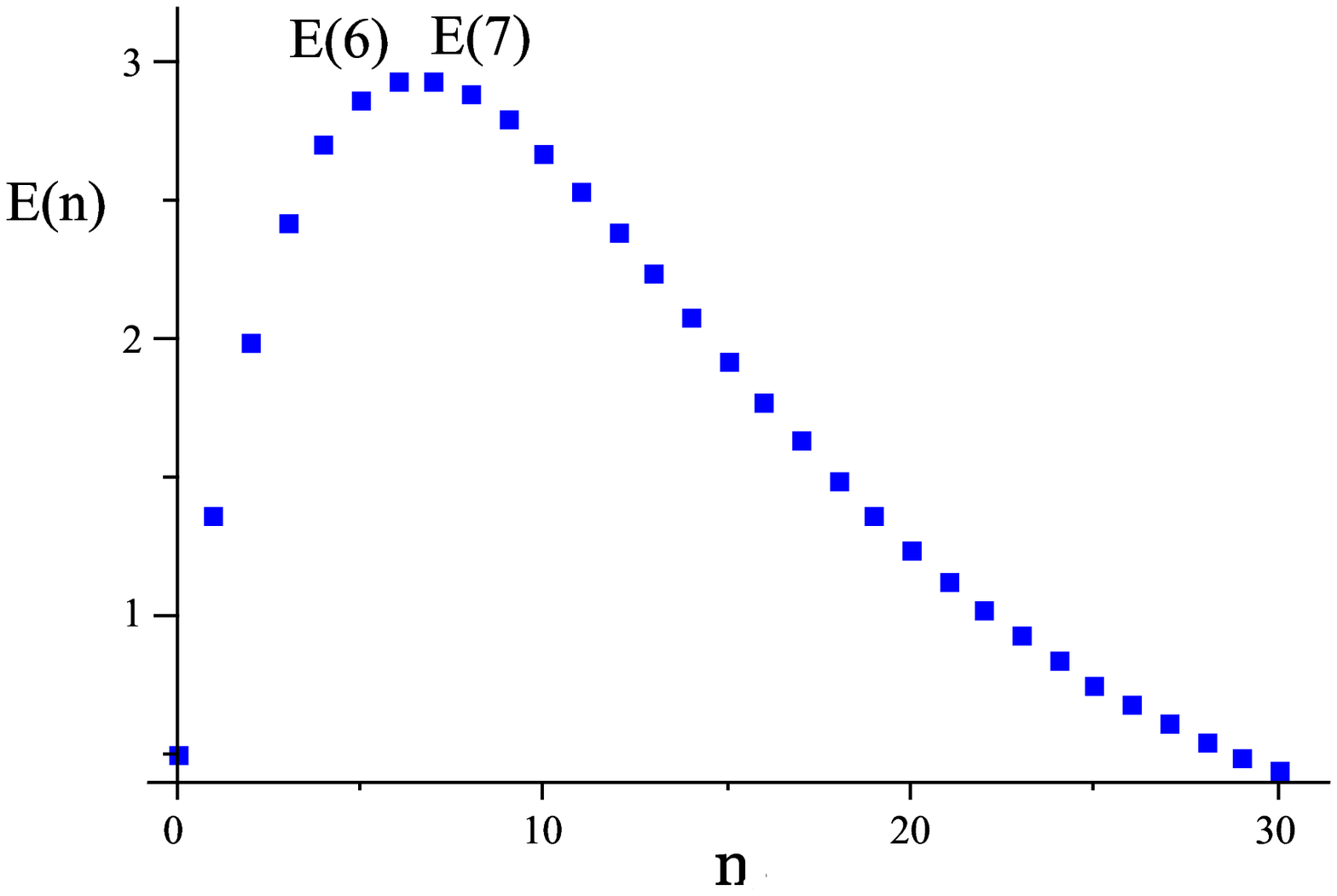}
\end{center}
%%%%%%%%%%%%%%%%%%%%%%%%%%%%%%%% figs. 1 %%%%%%%%%%%%%%%%%%%%%%%%%%%%
%\vspace{1mm}
%\vspace{1mm}

\vskip-3mm\noindent{\footnotesize Fig.~1. Spectrum of a
$q$-oscillator (8) at fixed $q=\sqrt{6/8}$. Observe the degeneracy
$E_{6}=E_{7}$ } \vskip15pt

\subsection*{2.2. \boldmath{$q$}-Oscillator with level
 degeneracy at real \boldmath{$q$}}

Although the AC $q$-oscillator does not allow any accidental
degeneracy, it was demonstrated in \cite{GR-1} that the
'accidental' degeneracy at real values of the $q$-parameter can
occur in the case of a $q$-oscillator called \cite{Odaka,Jagan}
the Tamm--Dancoff deformed oscillator. Its creation, destruction, and
number operators obey the following defining relation:
%8
\begin{equation}
b b^\dagger - q b^\dagger b = q^{N} , %\qquad
%  a a^\dagger - q^{-1} a^\dagger a = q^{N} ,
\end{equation}
%9
\begin{equation}
[{ N},b]=- b  ,     \qquad
[{ N},b^\dagger]= b^\dagger .   %\qquad \qquad \quad
\end{equation}

\noindent Taking the Hamiltonian of this $q$-oscillator as
%10
\begin{equation}
H  = \frac{\hslash\omega}{2} (b b^\dagger + b^\dagger b)
\end{equation}
and putting ${\hslash\omega}=1$, we consider its eigenvalues in the
states of the corresponding $q$-Fock space. With the vacuum state
$|0 \rangle$, the relevant relations are
%11
\begin{equation}
b |0 \rangle = 0 , \quad |n \rangle = \frac{(b^\dagger
)^n}{\sqrt{\{n \}_q!} } |0 \rangle , \quad \ N |n \rangle = n~|n
\rangle,
\end{equation}
where $\{n \}_q!=\{n \}_q\{n\!-\!1 \}_q ... \{2 \}_q\{1 \}_q$, \
$\{0\}!=1$, \ $\{1\}!=1$, and the $q$-bracket in this case being
%12
\begin{equation} \{X\}_q \equiv X q^{X-1} , \qquad  \quad
\{X\}_q \ \xrightarrow{q\to 1} X
\end{equation}
[compare it with (3)],
%13
\begin{equation}
b^\dagger  b = \{N\}_q , \quad
 b b^\dagger  = \{N+1\}_q  ,
\end{equation}
 and the operators  $b, \ b^\dagger$ act by the formulas
%14
\begin{equation}
b |n \rangle = \sqrt{\{n \}_q} |n\!-\!1 \rangle ,    \quad
b^\dagger |n \rangle = \sqrt{\{n\!+\!1 \}_q} \ |n\!+\!1 \rangle  .
\end{equation}
Note that, for any real $q\ge 0$, the operators $b$ and
$b^\dagger$ are adjoint to each other.

From (12)--(14), the spectrum $H|n\rangle = E_n |n\rangle$ of the
Hamiltonian reads
%15
\begin{equation}
E_n =  \frac12 \Bigl( (n+1) q^{n}  +  n q^{n-1} \Bigr) = \frac12
q^{n-1}\Bigl(q+n(1+q) \Bigr).
\end{equation}
At $q\to 1$, we recover $E_n = n + \frac12$; note also that $E_0 =
\frac12$ for any value of $q$.

If $q\ne 1$, the {\em spectrum is not uniformly spaced} (not
equidistant).    Moreover, if $q> 1$, the spacing $E_{n+1}-E_{n}$
gradually increases with growing $n$, so that $E_{n} \to \infty$ as
$n \to \infty$. However, more interesting possibilities %situation
arise when $q$ belongs to the interval $0<q<1$.

The energy spectrum given by expression (15) manifests some
sorts of degeneracies                                        \cite{GR-1},
with the strong dependence on the particular fixed value of $q$.
Let us consider relevant cases.

%\begin{center}
\vspace{15pt}
  \noindent \underline{\em Degeneracies $E_m = E_{m+1}$ and $E_m =
E_{m+2}$}
%\end{center}
\vspace{10pt}

\vspace{2mm} \noindent{\bf Proposition 3.} If the parameter $q$ is
fixed as $q=\sqrt{\frac{m}{m+2}}$, where $\ m\ge 1\ $, then the
following degeneracy of the energy levels does occur:
%16
\begin{equation}
E_m =E_{m+1}.
\end{equation}
Note that $m\!=\!0$ is excluded from (16) as the degeneracy
$E_0\!=\!E_1$ would require the (excluded) value $q\!=\!0$.

\vspace{2mm}
%%%%%%%%%%%%%%%%%%%%%
{\bf Proposition 4.} Let the parameter $q$ be fixed as
%17
\begin{equation}
q=\frac{1 + \sqrt{4 m^2 + 12 m + 1}}{2 ( m + 3)}  \quad \quad
     {\rm with}  \quad \quad   \   m\ge 0\ .
\end{equation}
Then the following degeneracy of $E_m$ does occur:
%18
\begin{equation}
E_m =E_{m+2}.
\end{equation}

For illustration, we show the particular cases
$E_6=E_{7}$ and $E_4=E_6$ of (16) and (18) in Figs.~1 and 2.
%%%%%

%\begin{center}
\vspace{15pt}
   \noindent\underline{\em Degeneracy of the type $E_0 = E_m$}
\vspace{15pt}
%\end{center}

\noindent One more type of degeneracy, $E_0 = E_m$, was also
pointed out in \cite{GR-1}, see the next proposition.

\end{multicols}
\begin{multicols}{2}

%\vspace{-2mm}
%%%%%%%%%%%%%%%%%%%%%%%%%%%%%%%%  figs. 2  %%%%%%%%%%%%%%%%%%%%%%%%%%%%
\begin{center}\noindent
\includegraphics[angle=0, width=8.4cm]
%%%{e46.eps}
{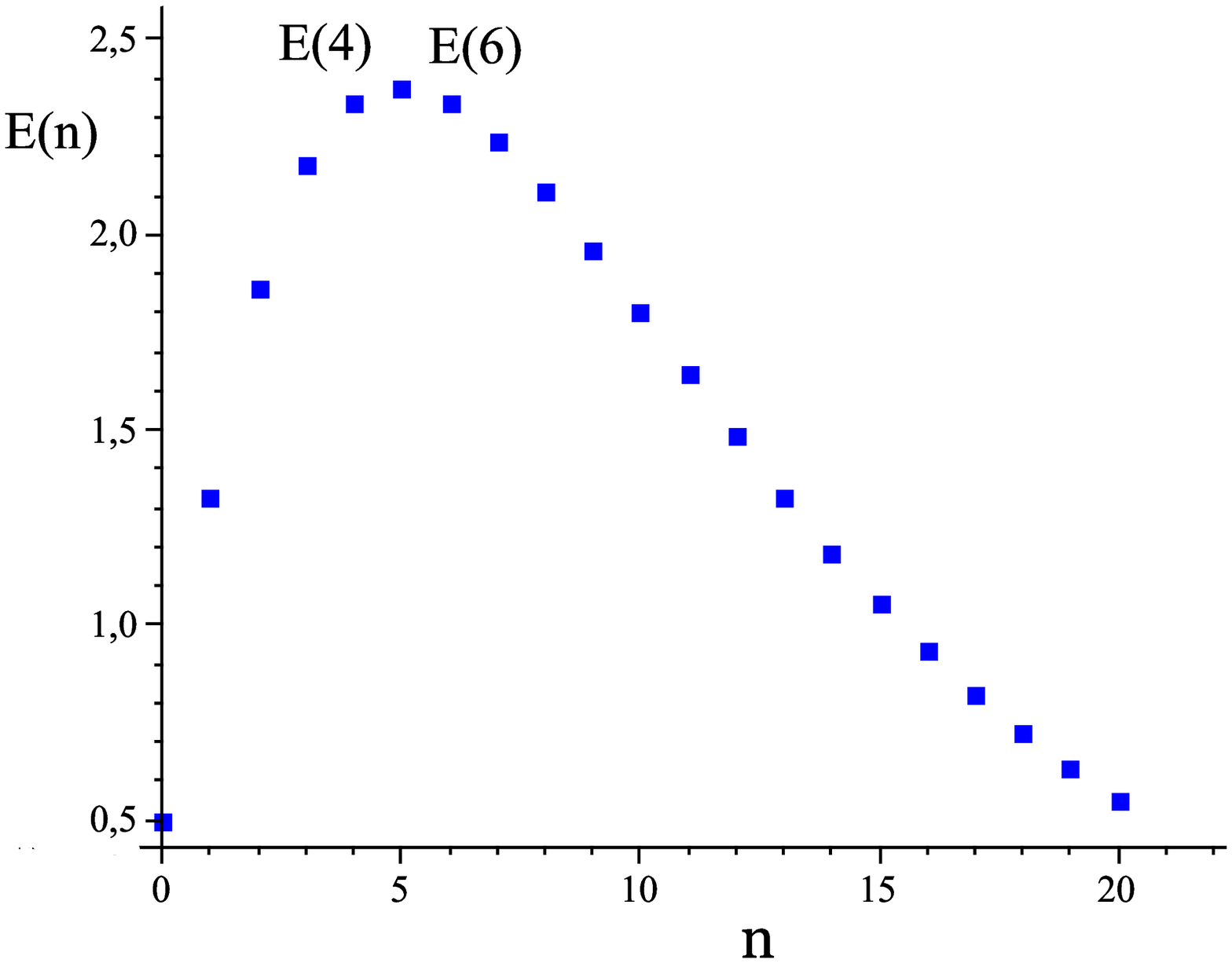}
%\vspace{-6mm}
\end{center}
%\vspace{1mm}

\vskip-3mm \noindent{\footnotesize Fig.~2. Spectrum of a
$q$-oscillator (8) at fixed $q=\frac{1+\sqrt{113}}{14}$.  Observe
the degeneracy $E_{4}=E_{6}$ } \vskip15pt

\noindent

%%%%%%%%%%%%%%%%%%%%%%%%%%%%%%%% figs. 3 %%%%%%%%%%%%%%%%%%%%%%%%%%%%
\begin{center}\noindent
\includegraphics[angle=0, width=7.7cm]
%%%{e04.eps}
{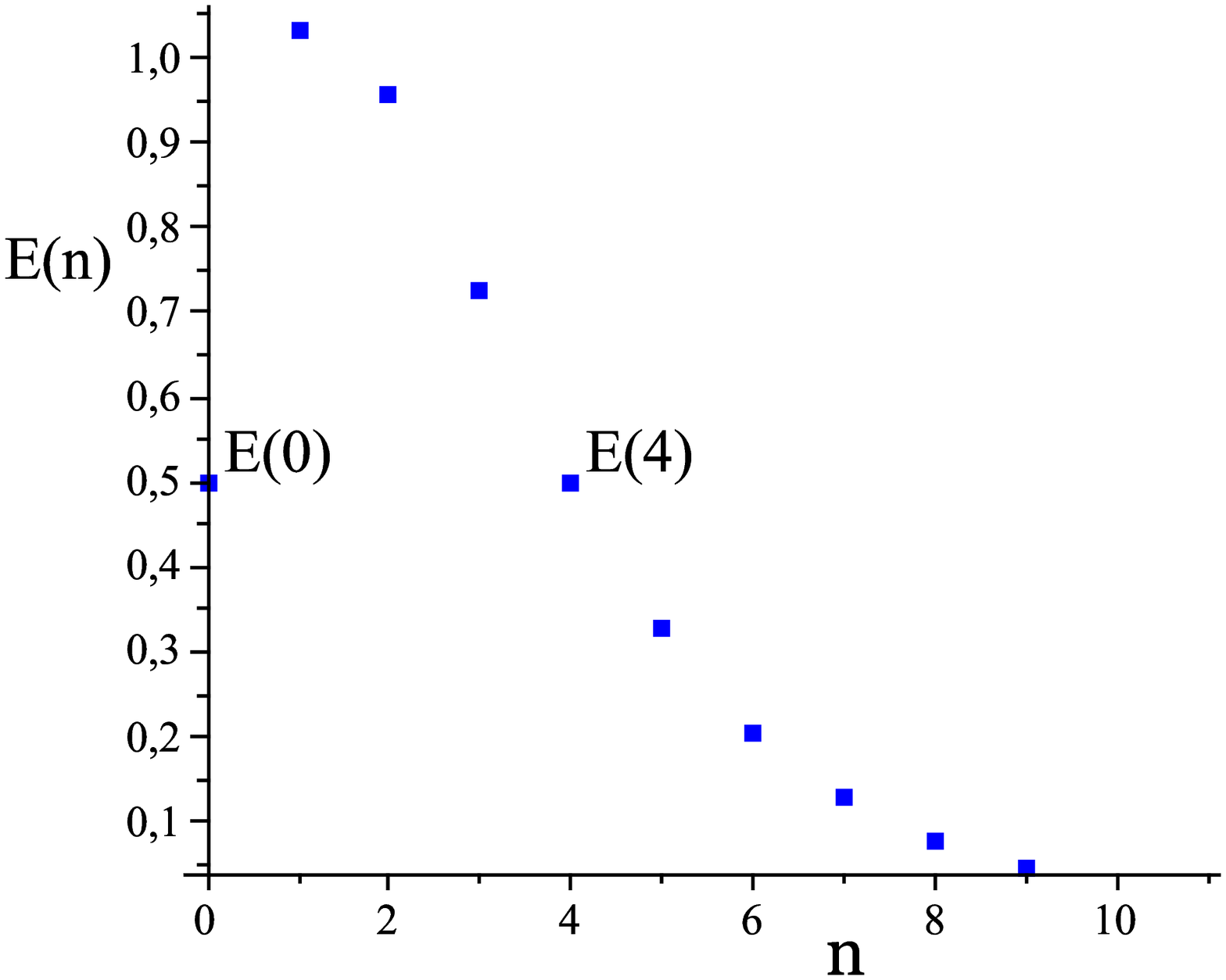}
%\vspace{-6mm}
    \end{center}
\vspace{-2mm}

\noindent{\footnotesize Fig.~3. Spectrum of the $q$-oscillator (8)
at fixed $q\simeq 0.5315645$.
Observe the degeneracy $E_{0}=E_{4}$ } %of the two energy levels:
%\vskip15pt       %\vspace{-6mm}
\end{multicols}
\begin{multicols}{2}

{\bf Proposition 5.}  For any integer $ m = 2, 3, 4, \ldots $,
there exists an appropriate $q_m = q(m)$ such that
%19
\begin{equation}
E_0 = E_m .
\end{equation}

The proof of this statement uses a graphical treatment as demonstrated
in \cite{GR-1}.

Some values of the $q$-parameter which provide degeneracy (19) are
listed in the Table. The first three values $q_2, q_3,$ and $q_4$
in the Table can be given in radicals, while, for $m\ge 5$ the
values $q_m$ are found approximately. Clearly, all the $q_m$ obey
the relation \mbox{$0 < q_m < 1$}.\looseness=1

 In Fig.~3, the particular case $m=4$ of Eq.~(19) is presented. The
latter degeneracy occurs, as seen from the Table, at $q\simeq
0.5315645$.

%\begin{center}
\vspace{15pt}
 \noindent  \underline{\em ``Accidental'' degeneracy of the general type $E_m = E_{m+k}$}
\vspace{5pt}
%\end{center}

\noindent As was mentioned in \cite{GR-1}, the diversity of
possible cases of two-fold degeneracy are given by the relation
$E_{m}=E_{m\!+\!k}$. The equation for the values $q=q(m,k)$ of
$q$-para-

\vskip2mm
\noindent{\footnotesize{\bf%
Some values \boldmath{$q_{m}$} that yield degenerace
\boldmath{$E_0=E_{m}$}}\vskip1mm \tabcolsep35.1pt

\noindent\begin{tabular}{l  l}
\hline%
\rule{0pt}{9pt}$m=2$ & ${q}_2= \frac13\simeq 0.333333$   \\
 $m=3$&
${q}_3 \simeq 0.45541$ \\ $m=4$ &${q}_4 \simeq 0.5315645$\\ $m=5$&
$q_5\simeq 0.585442$ \\
 $m=6$&$q_6\simeq 0.626225$\\
  $m=10$&$q_{10}\simeq 0.725405$\\
   $m=25$ &$q_{25}\simeq0.851675$\\
%$m=50$  &  $q_{50}\simeq 0.910968$   \\
$m=100$ &$q_{100}\simeq 0.948094$\\
%$m=200$  &  $q_{200}\simeq 0.9704016$   \\
$m=400$ &$q_{400}\simeq 0.983404$\\
 \hline
\end{tabular}}

\vspace{10mm}

\noindent meter responsible for such degeneracies looks as
%20
\[
 \ (m+k+1)q^{m+k} +
(m+k)q^{m+k-1}-
\]
\[ -(m+1)q^{m}- m q^{m-1} = 0,
\]
 or
\begin{equation}
(m+k+1)~q^{k+1} + (m+k)~q^{k} - (m+1)~q - m = 0.
\end{equation}
For each pair ($m$, $m + k$), it can be proved that there exists such
real solution $q\!=\!q(m,k)$ of (20) that $0\!<\!q\!<\!1$.
    Let us comment on few low $k$ values.
Obviously, $k=1$ resp. $k=2$ correspond to the particular series of
degeneracies already considered, see (16) and (18) above. For the
next two cases, the equations to be solved are
%21
\begin{equation}
%\hspace{-1mm}
({k=3})  \hspace{4mm} q^{4} + \frac{m+3}{m+4}~q^{3} -
\frac{m+1}{m+4}~q - \frac{m}{m+4} = 0 ,
\end{equation}
%22
\begin{equation}
({k\!=\!4})\!  \hspace{3mm} q^{4}\! -\! \frac{1}{m+5}~q^{3}\! +\!
\frac{1}{m+5}~q^2 \!-\! \frac{1}{m+5}~q \!- \!\frac{m}{m+5}\! =\!
0,
\end{equation}
where, for $k=4$, we have taken into account that the fifth-degree
equation divides exactly by $q+1$.
 Note also that the root $q=-1$ exists in all the cases of
higher even $k$ in (20). Equations (21) and (22) can be
solved in radicals, which yields awkward expressions.
  Analogously to the above Table, a set of values $q=q(m,k)$ can be
found numerically and tabulated.
 Let us finally remark that the case $E_0=E_m$ (see Proposition 5 above) is
obviously covered by the most general situation, $E_m=E_{m+k}$.

\section{``Accidental'' degeneracies of \boldmath$q,p$\,-oscillators}

The following main part of our paper deals with the issue of
degeneracies for the two-parameter extended or $q,p$\,-deformed
oscillators defined \cite{Chakr-Jag} by the relations
%23
\begin{equation}
 A A^\dagger - q\ A^\dagger A = p^{\cal N}   ,  \quad \quad
 A A^\dagger - p\ A^\dagger A = q^{\cal N}  ,
\end{equation}
along with two relations involving $A^\dagger$, $A$ and ${\cal N}$
completely analogous to (2) and (9).

The pair of relations (23) is symmetric under $q \leftrightarrow
p$ and leads to the formulas
%24
\begin{equation}
A^\dagger A = [\![{\cal N}]\!]_{q,p} \ ,
                                       \quad
A A^\dagger = [\![{\cal N}+1]\!]_{q,p} \ ,
                   \quad
                   \end{equation}
  where the $q,p$\,-bracket is
  %25
\begin{equation}
[\![X]\!]_{q,p} \equiv \frac{ q^{X}-p^{X} }{ q-p } \ .
\end{equation}
Obviously, with $p=q^{-1}$, we are back to the BM case
\cite{Bied,Mcf} of $q$-oscillators and to the AC case [5] at
$p=1$.
 The other special case [8, 9] of TD deformed oscillators corresponds to
$p=q$.

Similarly to BM and TD $q$-oscillators, we take the Hamiltonian in
the form
%26
\begin{equation}
H  = \frac{1}{2} (A A^\dagger + A^\dagger A) .
\end{equation}
In the $q,p$\,-deformed Fock space for which $A |0 \rangle = 0$,
%27
\begin{equation}
|n \rangle = \frac{(A^\dagger )^n}{\sqrt{[\![ n]\!]_{q,p}!} } |0
\rangle  \ , \quad \quad N |n \rangle = n~|n \rangle \ ,
\end{equation}
 the creation/annihilation operators act by the formulas  %according to
%28
\begin{equation}
A \ |n \rangle \!=\! \sqrt{[\![n]\!]_{q,p}} \ |n-1 \rangle\ , ~~
A^\dagger \ |n \rangle\! =\! \sqrt{[\![n+1]\!]_{q,p}} \ |n+1
\rangle \ .
\end{equation}
The spectrum $H|n\rangle = E_n |n\rangle$ of the Hamiltonian reads
%29
\begin{equation}
E_n =  \frac12 \Bigl( [\![n+1]\!]_{q,p}  +  [\![n]\!]_{q,p} \Bigr).
\end{equation}
As $q,p\to 1$, \ $E_n = n + \frac12$. In addition, $E_0 = \frac12$
for any~$q,p$\,.

To study the degeneracy properties of $q,p$\,-oscillators, we
consider $q,p$\, as real parameters valued in the intervals
%30
\begin{equation}
0 \leq q\le 1 ,    \qquad  \qquad  0\le p\le 1,
\end{equation}
where the point (0,0) is excluded.

Now we go over to 'accidental' degeneracies and demonstrate the validity
of relevant statements.

%%%%%%%%%%%%%%%%%%%%%

%\begin{center}
\vspace{15pt}
  \noindent \underline{\em Degeneracy of the type $E_m=E_0$}
\vspace{15pt}
%\end{center}

%%%%%%%%%%%%%%%%%%%%%

%%%%%%%%%%%%%%%%%%%%%
\noindent{\bf Proposition 6.} There exists a continuum of pairs of
the values ($q, p$) or the equivalent continuum of points of the
curve $F_{m,0}(p,q)=0$, for which the degeneracy
%31
\begin{equation}
E_m - E_{0} = 0 \ , \qquad  \qquad  m=2, 3, 4,\ldots
\end{equation}
does hold.
  The curve is given by the equation
  %32
\begin{equation}
F_{m,0}(q,p)\equiv\sum^{m}_{r=0}p^{m-r}q^{r} +
\sum^{m-1}_{s=0}p^{m-1-s}q^{s} - 1 = 0 .
\end{equation}

To prove the statement, take account of Eqs. (29), (25) in Eq. (31).
Then, Eq. (32) obviously follows.
   This formula implies nothing but a certain implicit function
$p=f_{m,0}(q)$ which is continuous and monotonically decreases on
the $q$-interval in~(30).
To confirm this assertion, let us consider the derivative
%33
\[
\frac{d p}{d q} = f'_{m,0}(q) = - \frac{\partial F_{m,0}}{\partial
q} \left(\frac{\partial F_{m,0}}{\partial p}\right)^{-1} =
\]\vspace*{-4mm}
\begin{equation}
=  - \frac{\sum\limits_{r=1}^{m}r p^{m-r}q^{r-1} +
\sum\limits_{s=1}^{m-1}s
p^{m-1-s}q^{s-1}}{\sum\limits_{r=0}^{m-1}q^r (m\!-\!r) p^{m-1-r} +
\sum\limits_{s=0}^{m-2}q^{s}(m\!-\!1\!-\!s)p^{m\!-\!2\!-\!s}} \ .
%E_m - E_{m+1} = 0 , \qquad  \qquad  m \ge 1,   % \qquad  \qquad  0\le p\le 1 .
\end{equation}
One can prove the following two facts: 1) for none point $(p,q)$
obeying (30), the derivative $\frac{\partial F_{m,0}}{\partial p}$
in the denominator of (33) turns into zero.
   2) For the intervals in (30), both
$\frac{\partial F_{m,0}}{\partial q}$ and $\frac{\partial
F_{m,0}}{\partial p}$ are positive. Then, the derivative $\frac{d
p}{d q}$ in Eq.(33) is always negative, and, thus, $f_{m,0}(q)$ is
a continuously decreasing implicit function represented by a flat
curve in the quadrant given by~(30).

 {\bf Remark 1.}
In fact, the values of  %the parameters
$p$ and $q$ from the admissible pairs $(p,q)$, i.e., those solving
Eq. (32), belong to the intervals $0<q<q_m$ and $0<p<p_m$ which,
since $p_m,q_m<1$, are smaller than the intervals in (30) (the value
$p_m=q_m$ that solve (32) at either $q=0$ or $p=0$ being put,
clearly depends on the fixed $m$).
Moreover, denoting
$q_\infty\equiv 1$ (since $q_m  \stackrel{m\to
\infty}{\longrightarrow} \ 1$), we have
%4
\begin{equation}
q_2 < q_3 < q_4< \ldots < q_{m-1} < q_m < \ldots < q_\infty=1.
\end{equation}

Now consider, for all $m\ge 2$,  the above derivative $f'_{m,0}(q)$
at the end points $(q,p)=(0,p_m)$ and $(q,p)=(q_m,0)$, where
$q_m=p_m$, as well as the derivative of each $f_{m,0}(q)$ at the
midpoint of the curve fixed by $p=q$.
 It is easy to see that $f'_{m,0}(q)\vert_{q=p}=-1$ for any $m$,
whereas at the both endpoints the derivatives are negative and
such that
  \[
f'_{m,0}(q)\vert_{q=q_m, p=0} <  -1 <  f'_{m,0}(q)\vert_{q=0,
p=p_m}\ < 0 .
\]
As a result, with $q$ growing from zero to $q_m$,
            %as $q$ runs from zero to one,
the derivative $f'_{m,0}(q)$ is always negative and continuously
decreases from $f'_{m,0}(q)\vert_{q=0, p=p_m}$ through $-1$ to
$f'_{m,0}(q)\vert_{q=q_m, p=0}$ .

\vspace{2mm}
%%%%%%%%%%%%%%%%%%%%%
{\bf Example 1.} Let $m=2$. In this case, the relation   %equating
\[
F_{2,0}(q,p)=p^2+pq+q^2 +p+q-1 = 0
\]
yields the function (explicit for this case only)
%35
\begin{equation}
  p=f_{2,0}(q)=\frac{-1-q+\sqrt{(1+q)(1-3q)+4}}{2}  %p^2+pq+q^2 +p+q-1
\end{equation}
 which monotonically decreases for $0\le q \le q_2$, where
\begin{equation}
q_2=(\sqrt5 -1)/2, ~~~~p_2=q_2.
\end{equation}
Then, with account of (33) and (36), we have
\[
f'_{2,0}(q)= - \frac{p+2q+1}{2p+q+1}=
              \begin{cases}-\frac{p_2+1}{2p_2+1}\simeq-0.7236, & q=0 ;
              \\[1mm]
                                                        \ \  - 1 ,           &  p=q ;  \\[1mm]
                                                       -\frac{2q_2+1}{q_2+1}\simeq-1.382, & p=0 . \cr
              \end{cases}
\]
Figure 4 illustrates this case (and also the cases $m=4$ and 7).

%%%%%%%%%%%%%%%%%%%%%
{\bf Remark 2.} The equations, from which the values $q_m$ are
deduced (see Remark 1), can be presented in the form $
q+1=\frac{1}{q}$  \   for $m=2$, \ $  q+1=\frac{1}{q^2}$  \ for $
m=3$,  $\ldots \ , \   q+1=\frac{1}{q^{m-1}}$ \ {\rm for any} $m$.
Such a form is convenient for applying the graphical treatment.
 From these equalities, the above inequalities (34) become more obvious.
%%%%%%%%%%%%%%%%%%%%%%%%%%%%

%\begin{center}
\vspace{15pt}
 \noindent  \underline{\em Degeneracy of the type $E_{m+1}=E_m$}
\vspace{15pt}
%\end{center}

%%%%%%%%%%%%%%%%%%%%%%%%%%%%%%%%%%%%%%%%%%

%%%%%%%%%%%%%%%%%%%%%%%%%%%%%%%%%%%%%%%%%%
\noindent{\bf Proposition 7.} There exists a continuous curve
$F_{m+1, m}(q,p)=0$ given by the continuum of pairs ($q,p$\,), for
which the degeneracy
%37
\begin{equation}
E_{m+1} - E_m = 0\ , ~~~~~  m \ge 1\ ,
\end{equation}
does hold. The equation for this curve is
%38
\begin{equation}
F_{m+1,m}(q,p)\equiv \!\sum^{m+1}_{r=0}\!p^{m+1-r}q^{r} -
\!\!\sum^{m-1}_{s=0}\!p^{m-1-s}q^{s} = 0 .
\end{equation}

%%%%%%%%%%%%%%%%%%%%%%%%%%%%%%%% figs. 4 %%%%%%%%%%%%%%%%%%%%%%%%%%%%
\begin{center}\noindent
\includegraphics[width=8.0cm]
%%%{e_0247.eps}
{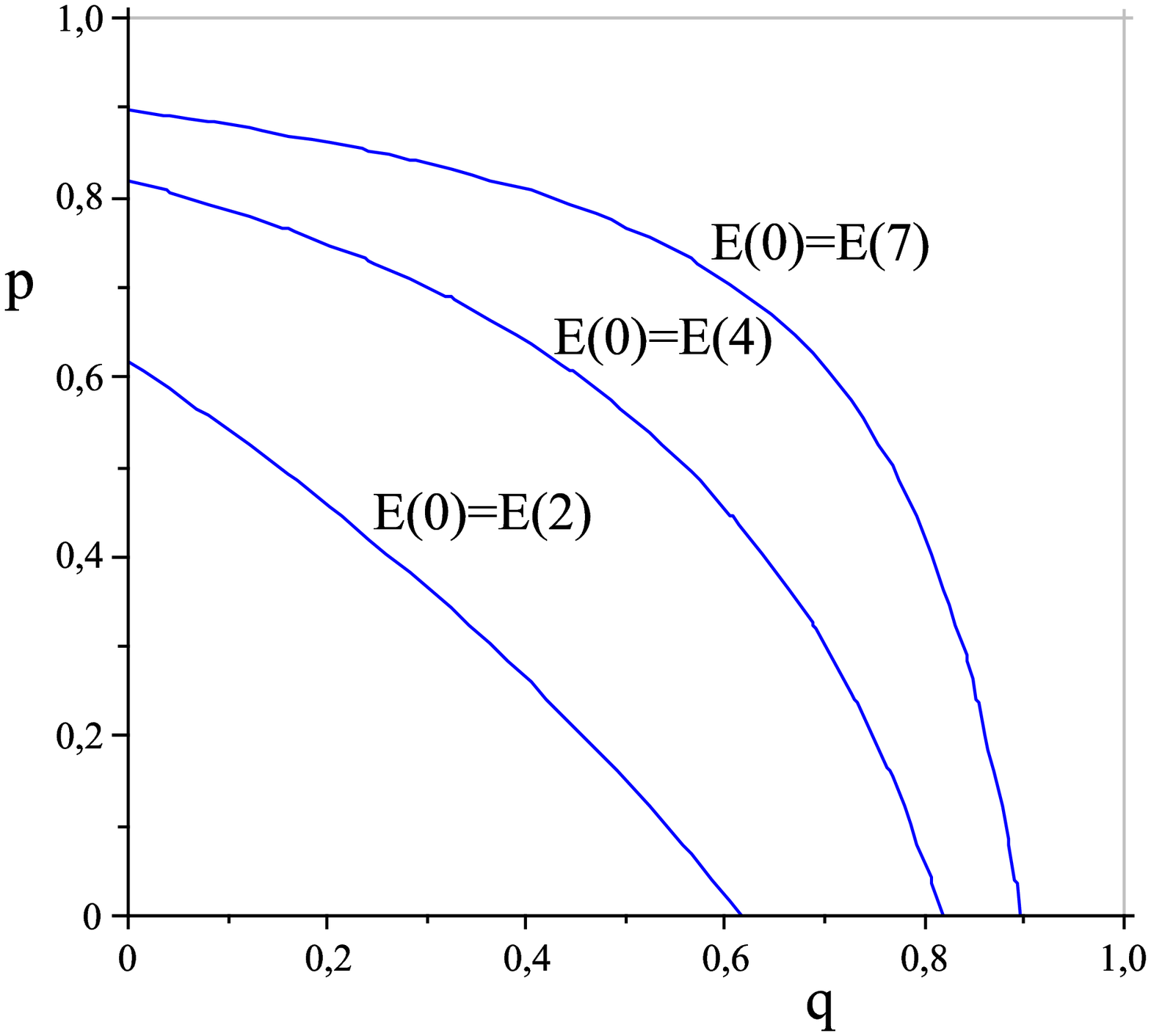}
%\vspace{-6mm}
\end{center}
%\vspace{-3mm}

\vskip-3mm \noindent{\footnotesize Fig.~4. Three cases of pairwise
degeneracies: $E_{0}=E_{2}$, $E_{0}=E_{4}$ and $E_{0}=E_{7}$ in
the energy spectrum (29) of a $q,p$\,-oscillator. The
corresponding curves are given by (31)--(32)} \vskip15pt
%\end{center}
%%%%%%%%%%%%%%%%%%%%%%%%%%%%%%%% figs. 4 %%%%%%%%%%%%%%%%%%%%%%%%%%%%

\noindent

In order to prove the statement, we substitute formula (29) in
Eq. (37), and Eq. (38) readily follows. Clearly, this
equation implies the continuous implicit function $p\!=\!f_{m+1,m}(q)$.
  To prove that this implicit function monotonically decreases
on the $q$-interval in (30), we examine the derivative
%39
\begin{equation}
\frac{d p}{d q} = f'_{m+1,m}(q) = - \frac{\frac{
\partial}{\partial q} F_{m+1,m}(q,p) }{ \frac{\partial}{
\partial p} F_{m+1,m}(q,p)
 },
\end{equation}
where
%40
\[
\frac{ \partial}{\partial q} F_{m+1, m}(q,p)=
\]\vskip-5mm
\begin{equation}
=\sum\limits_{r=1}^{m+1}r p^{m+1-r}q^{r-1} -
\sum\limits_{s=1}^{m-1}s p^{m-1-s}q^{s-1} ,
\end{equation}\vskip-5mm
%41
\[
\frac{ \partial}{ \partial p} F_{m+1, m}(q,p)=
\]\vskip-5mm
\begin{equation}
=\sum\limits_{r=0}^{m} (m\!+\!1\!-\!r)q^r p^{m-r}\! -\!\!
\sum\limits_{s=0}^{m-2}(m\!-\!1\!-\!s)p^{m-2-s}q^{s} .
\end{equation}
Obviously, the both partial derivatives are continuous (polynomial)
functions of two variables. The derivative in (41) should be nonzero
for each point of the flat curve

%%%%%%%%%%%%%%%%%%%%%%%%%%%%%%%% figs. 5 %%%%%%%%%%%%%%%%%%%%%%%%%%%%
\begin{center}
\noindent
\includegraphics[angle=0, width=8.0cm]
%%%{E12-45.eps}
{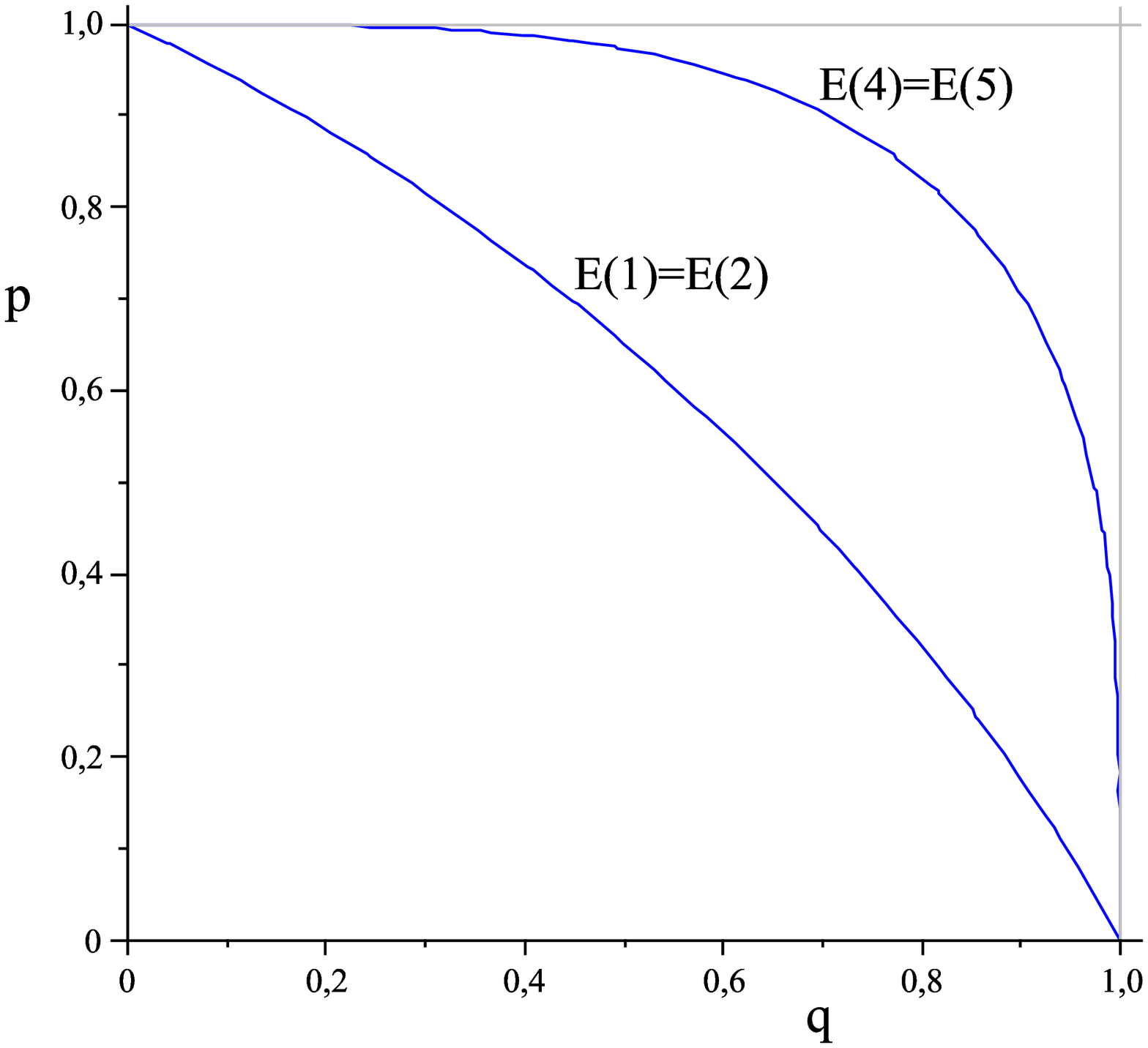}
 %\vspace{-1mm}
\end{center}

\vspace{-4mm} \noindent{\footnotesize Fig.~5. Two cases of
pairwise degeneracies, $E_{1}=E_{2}$ and $E_{4}=E_{5}$, in the
energy spectrum (29) of a $q,p$\,-oscillator} \vskip15pt
%\end{center}
%%%%%%%%%%%%%%%%%%%%%%%%%%%%%%%% figs. 5 %%%%%%%%%%%%%%%%%%%%%%%%%%%%

\noindent given by (38). To check this, we consider the set of
zeros of (41), that is, the set of pairs $(q_0,p_0)$ which solve
%42
\begin{equation}
\frac{ \partial}{ \partial p} F_{m+1, m}(q,p)=0 .
\end{equation}
One can show, for generic $m$, that such solutions $(q_0,p_0)$ form
a set of points none of which belongs to curve (38). Let us see
this in the particular cases of $m=1,2,3$.

For $m=1$ from (38)--(41), we have %the curve
\[
F_{2,1}\equiv p^2+pq+q^2\!-\!1=0 , \quad \frac{d p}{d q} = -
\frac{2q+p}{2p+q}<0 .
\]
Inserting $p+q=\frac1p(1-q^2)$ drawn from the equation of the
curve $F_{2,1}=0$ (or $E_2-E_1=0$) in the denominator of $\frac{d
p}{d q}$, we find that {\it for the points of the curve} the
denominator is $p+\frac1p(1-q^2)=\frac{p^2+1-q^2}{p}$ which is
strictly positive. Note also that $2p+q>0$ for all $q,p$\, from
(30).
%%%%%%%%%%%%%%%%%%%

For $m=2$, relations (38)--(41) yield %the curve
\[
 F_{3,2}\equiv p^3+p^2q+pq^2+q^3\!-p-q=0 ,
\]\vskip-5mm
\begin{equation}
\frac{d p}{d q} = - \frac{p^2+2qp+3q^2-1}{q^2+2pq+3p^2-1}<0 .
\end{equation}
Inserting $p^2+pq+q^2-1=\frac{q}{p}(1-q^2)$ drawn from the equation
of the curve $F_{3,2}=0$ (or $E_3-E_2=0$) in the denominator of
$\frac{d p}{d q}$ in (43), we find that {\it for the points of the
curve} the denominator is $\frac{q}{p}(1-q^2)+pq+2p^2$. That is, it is
always strictly positive (never turns into zero). Since the same
conclusion about strict positivity can be deduced for the numerator
in (43), the overall negative sign of the derivative $\frac{d p}{d
q}$ in (43) then follows.
%%%%%%%%%%%%%%%%%%%

For $m=3$, relations (38)--(41) yield %the curve
%44
\[
 F_{4,3}\equiv p^4+p^3q+p^2q^2+pq^3+q^4-p^2-pq-q^2=0,
\]\vskip-5mm
\begin{equation}
\frac{d p}{d q} = -
\frac{p^3+2qp^2+3q^2p+4q^3-p-2q}{q^3+2pq^2+3p^2q+4p^3-q-2p}<0.
\end{equation}
From the above equation $F_{4,3}=0$ for the curve of degeneracy
$E_4=E_3$, we draw $p^3+qp^2+q^2p+q^3-p-q=\frac{q^2}{p}(1-q^2)$
and insert it in the denominator of $\frac{d p}{d q}$ to get: {\it
for the points of the curve}, the denominator is
$\frac{q}{p}(1-q^2)+p(q^2+2pq+3p^2-1)$.
The latter is always      %
positive\footnote{Note that the polynomial in the second
parenthesis is identical to the denominator in (43) and, as argued
there, is strictly positive.} never turning into zero except for
the single point $(1,0)$. Since the same conclusion about strict
positivity can be deduced for the numerator in (44), the overall
negative sign of the derivative $\frac{d p}{d q}$ is confirmed.

So, for the cases of $m=1, 2, 3$, we have demonstrated that the
above implicit function $p\!=\!f_{m+1,m}(q)$ is a {\em continuous
monotonically decreasing} one. The proof can be extended to higher
values of $m$ and also to arbitrary $m$. In Fig.~5, the two
particular (different) degeneracy cases $E_{2}-E_{1}=0$ and
$E_{5}-E_{4}=0$ are shown.

\vspace{2mm}
 {\bf Remark 3.} The case $m=1$ of (38) (i.e. $E_1=E_2$)
differs from all other cases $m\ge 2$ since, at the end points
$(q,p)=(0,1)$ and $(q,p)=(1,0)$, the above derivative $f'_{m+1,
m}(q)$ has, for the $m=1$ case, the values differing from the rest
$m\geq 2$ cases. Namely, $f'_{2,1}(q)\vert_{q=0}=-\frac12$ and
 $f'_{2,1}(q)\vert_{q=1}=-2$. This implies that, as $q$ runs
from zero to one, the derivative $f'_{2,1}(q)$ continuously
changes from $-\frac12$ to $-2$. On the other hand, for all $m\ge
2$, we have $f'_{m+1,m}(q)\vert_{q=0}=0$ and $f'_{m+1,m}(q)
\xrightarrow{q\to 1} - \infty$, i.e., $f'_{m+1,m}(q)$ continuously
decreases from $0$ to $-\infty$ as $q$ grows from zero to one.
 The distinction of
$m=1$ (i.e. $E_{2}=E_{1}$) case from all other $m\ge 2$ cases (e.g.,
$E_{5}=E_{4}$) is clearly seen in Fig.~5.

 Let us emphasize that, contrary to the distinction just
discussed in Remark 3, all~ the~ degeneracy~ curves (38) of
$E_{m+1}-E_{m}=0$, with $m=1,2,\ldots$,  share the same value of
the derivative at their midpoints given by $p=q$:
$f'_{m+1,m}(q)\vert_{q=p}=-1$ (note also its coincidence with the
value $f'_{m,0}(q)\vert_{q=p}=-1$ mentioned in the paragraph
immediately after Eq.(34)). Clearly, this is rooted in the
$q\leftrightarrow p$ symmetry of the energy function [see (29) and
(25)] inherited by curves (38) and (32).

{\bf Remark 4.} Recall that the one-parameter deformed Tamm--Dancoff
oscillator, which stems from the $q,p$\,-oscillator if   %by setting
$p=q$, possesses double degeneracy \cite{GR-1}
of energy levels
$E_{m_1}=E_{m_2}$ at a certain value of the    %deformation
parameter $q$.
  %On the other hand, as shown
In the present paper, the two-parameter $q,p$\,-oscillator was
shown to possess the same type of degeneracy, $E_{m_1}=E_{m_2}$,
for the appropriate (continuum of) pairs $(q,p)$, where
 $q,p\!\in\!(0,1]$.
    This gives a hint of how is it possible to obtain, besides
the TD, numerous other $q$-deformed oscillators with a similar
property of double (pairwise) degeneracy of energy levels [15].
For such a degeneracy to occur in the chosen pair
$E_{m_1}=E_{m_2}$, it is necessary that the curve (in
$q,p$\,-plane) of the relation $p=f(q)$ generating the particular
$q$-oscillator intersects the
curve of degeneracy $E_{m_1}-E_{m_2}=0$ at least once. %For clarity,
This is displayed %we illustrate this
in Fig.~6 for a sample relation $p=q^5$ which crosses the indicated
%arbitrarily chosen
degeneracy curves $E_3-E_0=0$ and $E_5-E_4=0$.

It is clearly seen from Fig.~6 that the non-standard
$q$-oscillator inferred by substituting in (23)--(25) and (29) the
relation $p=q^5$ does possess the degeneracy $E_3-E_0=0$ at a
definite value of $q$ and the
 degeneracy $E_5-E_4=0$ at a distinct value of $q$.
 %\vspace{0.5mm}
Details of this approach with many particular cases are given in
[15].

\section{Conclusions and Outlook}

The study of deformed oscillators demonstrates that, due to
modified commutation relations, such oscillators possess
nontrivial properties very different from those of the standard
quantum oscillator. In our papers [7,15,16], we studied the
unusual property of accidental two-fold or double two-fold energy
level degeneracies of definite one-parameter deformed oscillators.
The present paper deals with the degeneracy of energy levels of
two-parameter deformed $q,p$\,-oscillators.

After recalling the special degeneracies occurring for the
Biedenharn--Macfarlane $q$-oscillator at $q$ being some roots of
unity, we placed a sketch of the 'accidental' double degeneracy
 properties \cite{GR-1}
of energy levels of the Tamm--Dancoff deformed oscillator. The
peculiarity of the latter consists in the fact that, for each pair
$E_{m+k}=E_{m}$ of energy levels, there exists a special {\bf real}
value of the $q$-parameter which provides their degeneracy.

In the main part of the paper, we have examined the ability of the
two-parameter $q,p$\,-oscillators to have pairwise energy level
degeneracies. As is shown, the $q,p$\,-oscillator possesses the
two-fold (pairwise) degeneracy of a definite type, i.e., within
some pair $E_{m_1}=E_{m_2}$, at the corresponding values $(q,p)$
from a continual set identical to the curve of $E_{m_2}-E_{m_1}=0$
in the $q,p$\,-plane.

%%%%%%%%%%%%%%%%%%%%%%%%%%%%%%%% figs. 6 %%%%%%%%%%%%%%%%%%%%%%%%%%%%
\begin{center}
\noindent
\includegraphics[width=7.2cm] %\textwidth]
%%%{0345E.eps}
{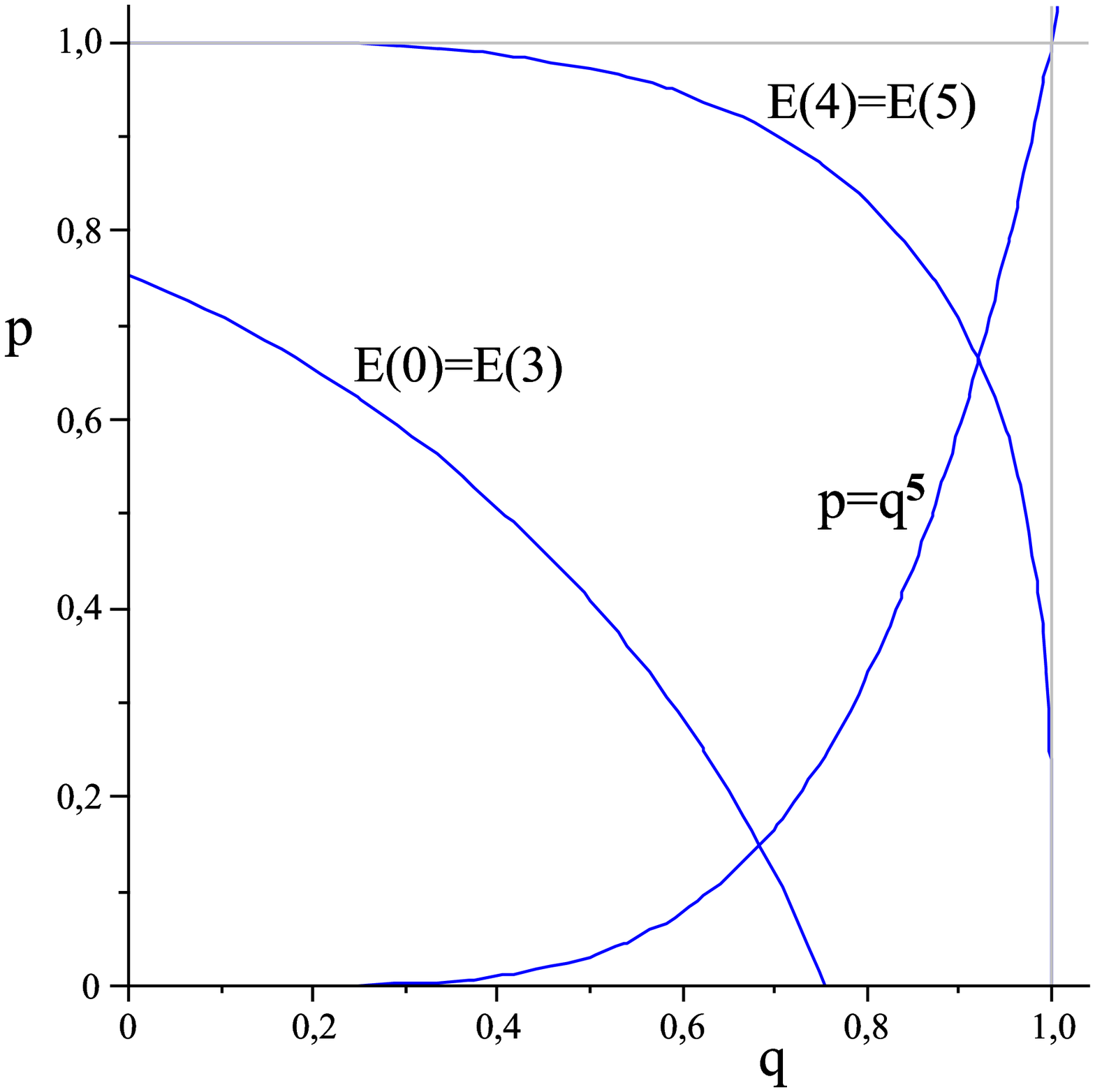}
\end{center}

\vspace{-4mm}
 \noindent {\footnotesize Fig.~6. Curve  %line of the relation
$p=q^5$ yielding a respective %that leads to definite
$q$-oscillator crosses the %pairwise
degeneracy curves $E_{0}=E_{3}$ and $E_{4}=E_{5}$ at different
values of $q$} \vskip15pt
%%%%%%%%%%%%%%%%%%%%%%%%%%%%%%%% figs. 6 %%%%%%%%%%%%%%%%%%%%%%%%%%%%

\noindent

What is important, the pairwise degeneracy of the energy levels
of $q,p$\,-oscillators observed at certain values of $q$ and $p$
is
``accidental'' %one
(as it occurs without any underlying symmetry) and involves a single
fixed pair of levels.

 Let us also remark that the degeneracy in $q,p$\,-oscillators shown
 in this paper is not in conflict, as it was already
 commented in [16],
with the well-known  ``no-go''  theorem [17, 18]
 about the absence, in one dimension, of
degenerate discrete states in any standard quantum-mechanical
system. Indeed, the $q,p$\,-oscillators analyzed in our paper go
beyond the scope of customary systems of traditional quantum
mechanics, due to such more general nontrivial features (see,
e.g., \cite{Mizrahi}) as the non-constant position-dependent mass
given by an inertia function, the complicated interaction
depending on both the position and the momentum, etc.

On the base of the considered (seemingly, unnoticed earlier)
important peculiarity of the $q,p$\,-oscillators, we can infer a
plenty of new nonlinear one-parameter deformed oscillators which
exhibit nontrivial and unusual degeneracy properties (diverse
patterns of levels degeneracies, including rather complicated
ones). As some step already made in this direction, let us quote
the paper [16], where a number of $p$-oscillators is presented
exhibiting a rather nontrivial pattern of two-fold double
degeneracies (two pairwise degeneracies within each of two fixed
pairs of energy levels, e.g., $E_1=E_2$ and $E_3=E_4$).

It is worth to mention again the fact of the applicability
\cite{SIGMA} of $q,p$\,-oscillators ($q,p$\,-bosons) in the
context of the efficiency of a description of the observed
non-Bose properties of the two- and multi-pion \mbox{(-kaon)}
correlations in the experiments on relativistic heavy-ion
collisions like that of the other types, e.g., the one-parameter
BM-type $q$-oscillator and the $q$-Bose gas model [19, 20]. In
that context, it would be interesting to find some peculiarity (if
any) connected with the feature of 'accidental' double degeneracy
of $q,p$\,-oscillators considered in the present paper. The same
can be said about the usage (see \cite{SIGMA}) of TD
$q$-oscillators and the ``TD $q$-Bose gas'' model which is just
the one-parameter $p\!=\!q$ limit of the $q,p$\,-Bose gas model.

One may also hope that the unusual novel $q$-bosons (related with
non-standard $q$-oscillators treated in [7, 15, 16]) and possibly
some others will be useful in the study of explicitly solvable
problems, say, along the lines similar to those described in [21],
and also for diverse physical applications.
%\vspace{3mm}

Of course, it is desirable to give explicit and exhaustive proofs
of the pairwise degeneracy of the energy levels of
$q,p$\,-oscillators for more involved cases like $E_{m+2}=E_{m}$
and, also, for the most general case of degeneracy:
$E_{m+k_1}=E_{m+k_2}$, $k_1 \neq k_2$. This will be done in a
separate paper.

\vskip3mm

This work was partially supported by the Grant 14.01/016 of the
State Foundation of Fundamental Research of Ukraine and by the
Special Program of the Division of Physics and Astronomy of the NAS of
Ukraine.

\rezume{%
ДВОКРАТНЕ ВИПАДКОВЕ ВИРОДЖЕННЯ \\ЕНЕРГЕТИЧНИХ РІВНІВ  У МОДЕЛІ
$q,~p$-ОСЦИЛЯТОРА } {О.М. Гаврилик, А.П. Ребеш} {Показано, що
двопараметрично деформовані осцилятори з параметрами деформації
$q,p$\,, де $0<q,~p\le 1$, мають властивість ``випадкового''
двократного виродження енергетичних рівнів типу $E_m=E_{m+1}$ та
типу $E_0=E_{m}$ при відповідних значеннях $q$ і $p$. Коротко
обговорено також найбільш загальний випадок виродження
$E_{m+k}=E_m$, де $k\ge 1$ для $m\ge 1$ або $k\ge 2$ для $m=0$.}

%\rezume{%
% ДВУКРАТНОЕ СЛУЧАЙНОЕ ВЫРОЖДЕНИЕ ЭНЕРГЕТИЧЕСКИХ УРОВНЕЙ В МОДЕЛИ
% $q,p$\,\,-ОСЦИЛЛЯТОРА}
% { А.М. Гаврилик, А.П. Ребеш} {В работе показано, что
% двухпараметрически деформированные осцилляторы с параметрами
% деформации $q,p$\,, где $0<q,p\le 1$, обладают свойством "случайного"
% \ двукратного вырождения энергетических уровней класса
% $E_m=E_{m+1}$ и класса $E_0=E_{m}$ при определённых значениях $q$ и
% $p$. Рассмотрен также наиболее общий случай вырождения
% $E_{m+k}=E_m$, где $k\ge 1$ для $m\ge 1$ или $k\ge 2$ для $m=0$.}

\end{multicols}
\end{document}